# Radioactive contamination of BaF$_2$ crystal scintillator[a]

O.G. Polischuk[1,2,b], P. Belli[3], R. Bernabei[3,4], F. Cappella[2,5], V. Caracciolo[6], R. Cerulli[6], A. Di Marco[4], F.A. Danevich[1], A. Incicchitti[2], D.V. Poda[1], V.I. Tretyak[1]

[1]Institute for Nuclear Research, MSP 03680 Kyiv, Ukraine
[2]INFN, Sezione di Roma, I-00185 Rome, Italy
[3]INFN, Sezione di Roma "Tor Vergata", I-00133 Rome, Italy
[4]Dipartimento di Fisica, Università di Roma "Tor Vergata", I-00133 Rome, Italy
[5]Dipartimento di Fisica, Università di Roma "La Sapienza", I-00185 Rome, Italy
[6]INFN, Laboratori Nazionali del Gran Sasso, I-67100 Assergi (AQ), Italy

**Abstract.** Barium fluoride (BaF$_2$) crystal scintillators are promising detectors to search for double beta decay processes in $^{130}$Ba ($Q_{2\beta}$ = 2619(3) keV) and $^{132}$Ba ($Q_{2\beta}$ = 844(1) keV). The $^{130}$Ba isotope is of particular interest because of the indications on 2$\beta$ decay found in two geochemical experiments. The radioactive contamination of BaF$_2$ scintillation crystal with mass of 1.714 kg was measured over 113.4 hours in a low-background DAMA/R&D set-up deep underground (3600 m w.e.) at the Gran Sasso National Laboratories of INFN (LNGS, Italy). The half-life of $^{212}$Po (present in the crystal scintillator due to contamination by radium) was estimated as $T_{1/2}$ = 298.8 ± 0.8(stat.) ± 1.4(syst.) ns by analysis of the events pulse profiles.

## 1 Introduction

BaF$_2$ is a promising crystal scintillator for different applications (detection of high energy gamma rays [1] and neutrons [2]). The scintillation material is also widely used in medicine for positron emission tomography (PET) [3]. The compound looks a promising detector to search for double beta (2$\beta$) decay of barium [4]. Natural barium contains two potentially 2$\beta$ active isotopes, $^{130}$Ba ($Q_{2\beta}$ = 2618.7(2.6) keV) and $^{132}$Ba ($Q_{2\beta}$ = 844.0(1.1) keV) [5, 6]. The $^{130}$Ba isotope is of particular interest because of the reports on the observation in geochemical experiments of double electron capture with a half-life $T_{1/2}$ = (2.2 ± 0.5) × 10$^{21}$ yr [7] and $T_{1/2}$ = (6.0 ± 1.1) × 10$^{20}$ yr [8].

The high level of radioactive contamination of BaF$_2$ scintillation crystal by uranium and thorium is the main source of background of the detectors [4], however, this feature allows to use the detector for measurements of some short-lived isotopes in U/Th chains (e.g. of $^{212}$Po). Results of measurements of radioactive contamination of a large volume BaF$_2$ crystal scintillator are presented here. We have also derived a half-life value of $^{212}$Po from the data by using pulse-shape analysis of $^{212}$Bi – $^{212}$Po events.

## 2 Measurements, results and discussion

### 2.1 Experiments

Radioactive contamination of BaF$_2$ (3" × 3", 1.714 kg) was measured over 113.4 hours in a low-background DAMA/R&D set-up deep underground (3600 m w.e.) at the Gran Sasso National Laboratories of INFN (LNGS, Italy). The BaF$_2$ crystal scintillator was viewed through two light-guides (⌀3" × 100 mm) by two low radioactive 3" photomultipliers (PMT, ETL 9302FLA). The detector was surrounded by Cu bricks and sealed in a low radioactive air-tight Cu box continuously flushed by high purity nitrogen gas (stored deep underground for a long time) to avoid the presence of residual environmental radon. The Cu box was surrounded by a passive shield made of high purity Cu, 10 cm of thickness, 15 cm of low radioactive lead, 1.5 mm of cadmium and 4 to 10 cm of polyethylene/paraffin to reduce the external background. The shield was contained inside a Plexiglas box, also continuously flushed by high purity nitrogen gas. An event-by-event data acquisition system recorded amplitude and pulse profile of events at the sampling rate of 1 GSPS over a time window of 4000 ns.

---

The energy scale of the BaF$_2$ detector and its energy resolution in the range of interest have been determined by means of $^{22}$Na (511, 1275 keV), $^{137}$Cs (662 keV), $^{241}$Am (60 keV), $^{60}$Co (1173, 1333 keV), $^{133}$Ba (356 keV) and $^{228}$Th (239, 2615 keV) gamma sources. The energy resolution (full width at half maximum, FWHM) of 15.5% was obtained for 662 keV γ quanta of $^{137}$Cs, while for 511 and 1275 keV γ lines of $^{22}$Na source the energy resolution was 16.4% and 10.8%, respectively (see Fig. 1; in all figures, energy is given in gamma scale). The energy dependence of the energy resolution can be approximated as: FWHM (keV) = [397(54) + 15.6(3) × $E_\gamma$]$^{1/2}$, where $E_\gamma$ is energy of gamma quanta in keV.

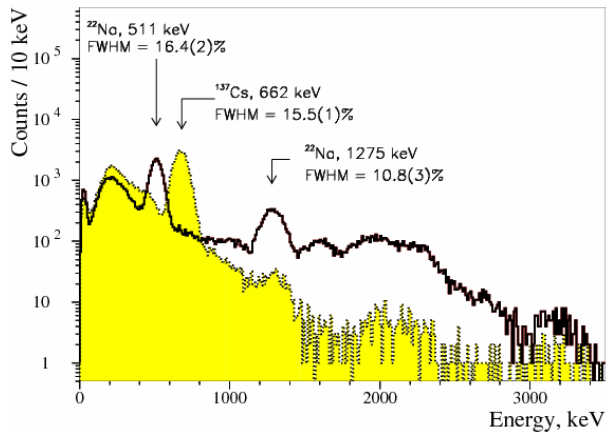

**Figure 1.** Energy spectra accumulated with BaF$_2$ detector with $^{137}$Cs and $^{22}$Na gamma sources.

### 2.2 Pulse-shape discrimination of α, β and Bi-Po events

The mean time method was used to separate alpha, gamma(beta) and Bi-Po events in the BaF$_2$ scintillator. The scatter plot of the mean time versus energy (see Fig. 2) demonstrates pulse-shape discrimination ability of the detector.

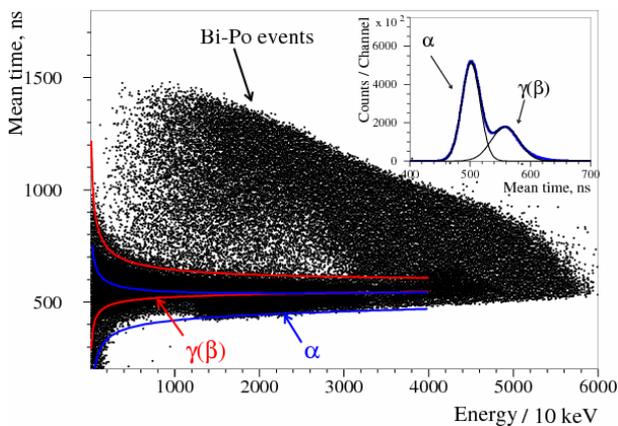

**Figure 2.** Scatter plot of the mean time versus energy accumulated by the BaF$_2$ scintillation detector over 113.4 hours. (Inset) The mean time spectrum in the energy interval 1200–3400 keV. The α and γ(β) events distributions are fitted by Gaussian functions (solid lines).

The background energy spectrum accumulated over 113.4 hours with the BaF$_2$ crystal scintillator is presented in Fig. 3. A substantial increase of the counting rate in the energy interval 1.2 – 3.4 MeV is due to α activity of $^{238}$U and $^{232}$Th daughters from uranium and thorium contamination of the BaF$_2$ crystal.

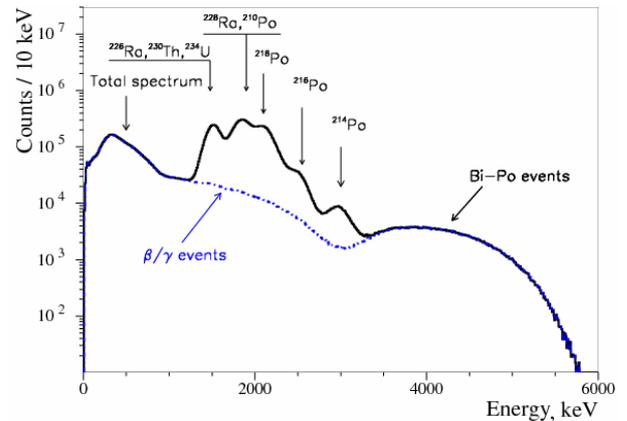

**Figure 3.** Background energy spectrum of the BaF$_2$ scintillator collected during 113.4 hours. The spectrum of β/γ events separated by the pulse-shape discrimination is shown by dotted line.

Energy spectrum of α events selected from the background data with the help of the pulse-shape discrimination is presented in Fig. 4. The spectrum was fitted by a model built of α peaks of $^{238}$U and $^{232}$Th and their daughters, assuming broken equilibrium of the chains. Therefore the activities of the following nuclides and sub-chains ($^{232}$Th, $^{228}$Th-$^{212}$Pb; $^{238}$U, $^{234}$U, $^{230}$Th, $^{226}$Ra-$^{214}$Po, $^{210}$Po; $^{235}$U, $^{231}$Pa, $^{227}$Ac-$^{211}$Bi) were taken as free parameters of the fit. Besides, values of the energy resolution of the detector to alpha particles and alpha/beta ratio (relative light output for α particles as compared with that for β particles (γ rays)) were also introduced into the fit as free parameters.

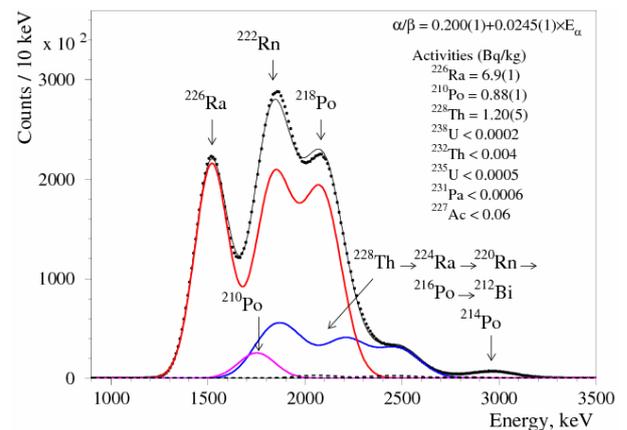

**Figure 4.** Energy spectrum of α events selected by the pulse-shape discrimination from the data of low-background measurements with the BaF$_2$ crystal over 113.4 hours (points). Fit of the data by the model built from α decays of $^{238}$U and $^{232}$Th daughters is shown by solid lines.

According to the fit, the α/β ratio for the BaF$_2$ scintillation detector depends on energy of α particles as α/β = 0.200(1) + 0.0245(1) × $E_\alpha$, where $E_\alpha$ is energy of α particles in MeV. The radioactive contamination of the

BaF$_2$ crystal obtained from the fit is presented in Table 1. One can conclude that the BaF$_2$ crystal is contaminated by radium.

**Table 1.** Radioactive contamination of the BaF$_2$ crystal.

| Chain | Source | Activity, Bq/kg |
|---|---|---|
| $^{232}$Th | $^{232}$Th | < 0.004 |
|  | $^{228}$Th | 1.20(5) |
| $^{238}$U | $^{238}$U | < 0.0002 |
|  | $^{226}$Ra | 6.9(1) |
|  | $^{210}$Pb | 0.88(1) |
| $^{235}$U | $^{235}$U | <0.0005 |
|  | $^{231}$Pa | < 0.0006 |
|  | $^{227}$Ac | < 0.06 |

## 2.3 Pulse-shape analysis of Bi-Po events

Events in the fast chains ($^{212}$Bi–$^{212}$Po from $^{232}$Th family and $^{214}$Bi–$^{214}$Po from $^{238}$U) can be selected by the mean time method, as one can see in Fig. 2. The time intervals between β events (of $^{212}$Bi or $^{214}$Bi) and subsequent α events (of $^{212}$Po or $^{214}$Po) were obtained by analysis of the pulse profiles. An example of the $^{212}$Bi→$^{212}$Po→$^{208}$Pb event in the BaF$_2$ scintillator is presented in Fig. 5.

An energy spectra of the first events (beta decay of $^{212}$Bi with $Q_\beta$ = 2254 keV and $^{214}$Bi with $Q_\beta$ = 3272 keV) and of the second events (alpha decay of $^{212}$Po with $Q_\alpha$ = 8954 keV and $^{214}$Po with $Q_\alpha$ = 7833 keV) selected from the Bi-Po pulse profiles are presented in Fig. 6.

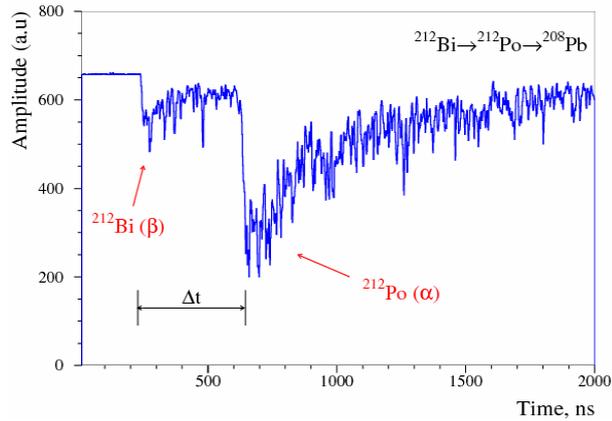

**Figure 5.** Example of the Bi-Po event in the BaF$_2$ scintillator.

# 3 Half-life of $^{212}$Po

A distribution of the time intervals selected by the pulse-shape analysis of the Bi-Po events is presented in Fig. 7. In order to suppress a contribution of the $^{214}$Bi–$^{214}$Po chain (with the half-life of $^{214}$Po $T_{1/2}$ = 164 μs), the energy of the second event was selected from the energy interval 3000–3800 keV (see lower part in Fig. 6). An energy threshold of 300 keV was chosen for the first events to decrease jitter of the event time determination. The time spectrum was fitted by sum of two exponential functions that represent the decays of $^{212}$Po and $^{214}$Po. The decay constant of the second exponent was bounded taking into account the table uncertainty of the $^{214}$Po half-life ± 2.0 μs [9].

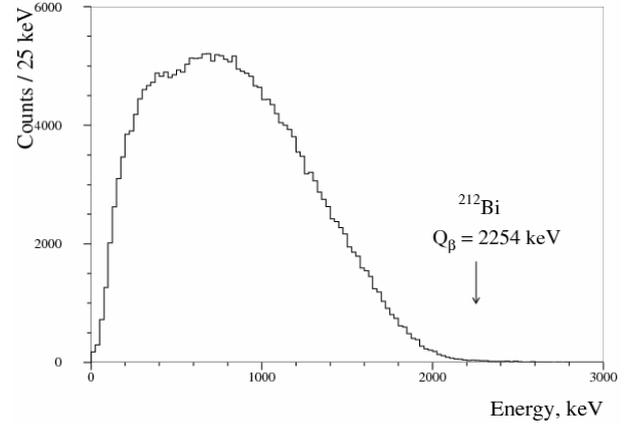

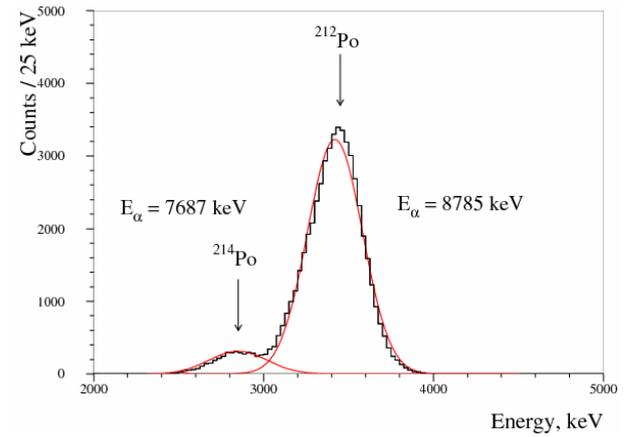

**Figure 6**. Energy spectra of the first (mainly $^{212}$Bi with $Q_\beta$ = 2254 keV; upper part) and of the second ($^{212}$Po, $Q_\alpha$ = 8954 keV; lower part) events selected by the pulse-shape analysis of the background data accumulated with the BaF$_2$ crystal scintillator over 113.4 hours.

The time spectrum was fitted by the chi-square method in different time intervals from 100 ns to 1550 ns. To estimate a systematic error of the half-life value, we have analysed three time distributions with bins 1, 2 and 3 ns per channel. Only the fits with $\chi^2$/n.d.f. values in the range of 0.92 – 1.15 (where n.d.f. is number of degrees of freedom) were taken into account in the analysis. The fits gave 45 values of $^{212}$Po half-life in the range of 297.4 – 299.8 ns with an average value $T_{1/2}$ = 298.8 ± 0.8(stat.) ± 1.4(syst.) ns. The obtained half-life is an excellent agreement with the table value 299 ± 2 ns [9], and in a reasonable agreement with the recent result of the Borexino collaboration 294.7 ± 0.6(stat.) ± 0.8(syst.) ns [10].

We have also estimated activity of $^{228}$Th from the Bi-Po analysis as 1.04(10) Bq/kg, which is in a reasonable agreement with the result obtained from the fit of the alpha spectrum presented in Fig. 6.

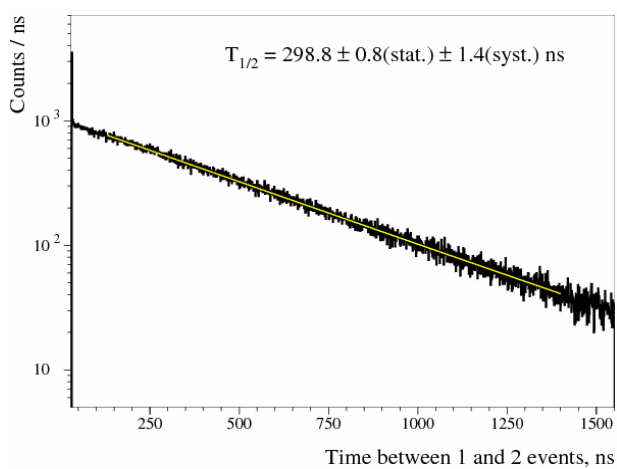

**Figure 7.** The time distribution for the fast sequence of β ($^{212}$Bi) and α ($^{212}$Po) decays selected by the pulse-shape analysis of Bi-Po events from the data accumulated with the BaF$_2$ scintillation detector over 113.4 hours. The obtained half-life $T_{1/2}$ = 298.8 ± 0.8(stat.) ± 1.4(syst.) ns is in good agreement with the table value 299 ± 2 ns [9].

## 4 Conclusions

The radioactive contamination of the BaF$_2$ crystal scintillator was estimated to be on the level of few Bq/kg for $^{226}$Ra and $^{228}$Th. Taking into account 3 orders of magnitude lower activity of $^{238}$U and $^{232}$Th (only limits <0.0002 Bq/kg for $^{238}$U and <0.004 Bq/kg for $^{232}$Th were obtained) and broken equilibrium in the chains, one can conclude that the BaF$_2$ crystal is contaminated by radium. The response of the BaF$_2$ crystal scintillator to α particles has been investigated in the wide energy interval and the capability of pulse-shape discrimination between α particles and γ quanta (electrons) has been demonstrated.

Analysis of the time intervals distribution between β and α decays in the fast Bi-Po chains allowed us to estimate half-life of $^{212}$Po as $T_{1/2}$ = 298.8 ± 0.8(stat.) ± 1.4(syst.) ns, which is in an agreement with the table value [9].

The contamination of the BaF$_2$ crystal by $^{226}$Ra and $^{228}$Th is the main problem in applications of this scintillator to search for double beta decay of barium. An R&D of methods to purify barium from radium traces is in progress at the Gran Sasso National Laboratories with an aim to develop radiopure BaF$_2$ crystal scintillators to search for double beta decay of $^{130,132}$Ba. Such a counting experiment is of particular interest, taking into account indications of two geochemical experiments on double beta decay of $^{130}$Ba.